\documentstyle[12pt,a4]{article}

\begin{document}

\title{
\Large{\bf On the classification of \\
spatially homogeneous 4D string backgrounds}}
\author{\normalsize
\bf  Nikolaos A. Batakis
\thanks{Permanent address: Dept of Physics, Univ of
Ioannina,
GR--45100}
\thanks{e-mail address: batakis@surya20.cern.ch}\\
\normalsize Theory Division, CERN,
     CH--1211 Geneva 23, Switzerland}

\date{}
\maketitle

\vspace{3cm}
\begin{abstract}
\begin{sloppypar}
\normalsize

A classification of all possible spatially homogeneous 4D string backgrounds
(HSBs)
has been obtained by appropriate ramification of
the existing nine Bianchi types of homogeneous 3D spaces. A total of $24^2=576$
HSBs which have been classified as distinct contains a subclass of 192 which
includes all possible
FRW models as well as those in which SO(3) isotropy is attained asymptotically.
A discussion of these results also aims to fascilitate the identification of
HSBs
which have already appeared in the literature. The basic physical perspective
of the
parameters of classification is outlined together with certain features
relating to deeper aspects of string theory.
\end{sloppypar}
\end{abstract}
\vspace{2cm}
\addtolength{\baselineskip}{.3\baselineskip}
\newpage
\section{Introduction}

Considerable interest is being currently focused on spatially
homogeneous \cite{1} but not necessarily isotropic 4D string backgrounds
\cite{2}-\cite{6}.
The class of such spacetimes, hereafter
recaled as HSBs,
contains a large number
of the known string backgrounds, which either descend from a
conformal field theory and higher-dimensional compactifications, or simply
satisfy the lowest-order string beta function eguations. At the limit of
complete isotropy,
taken mathematically or attained
dynamically, the class of HSBs contains all possible Friedmann-Robertson-Walker
or FRW-like
backgrounds. On the other hand, non-isotropic HSBs seem to provide the best
models
for the understanding of anisotropy and its potentially crucial impact on the
dynamics of the early universe \cite{6}, well before the attainment of the
observed state of isotropy. This is precisely the region where the
most fundamental cosmological problems arise and also where string
theory seems to have its best chance of being confronted with reality.
The class of HSBs admits a rather elegant mathematical treatment,
exploitable for calculations as well as transparency to the deeper aspects
of string theory.

It is not only desirable but also quite conceivable that most (if not all) HSBs
will be found.
With some progress in that direction already at hand \cite{2}-\cite{5},
we may already proceed
to establish a full classification scheme for all HSBs.
By that we do not just mean the existing Bianchi classification of spatially
homogeneous spacetimes, already exploited in the conventional formulation of
general relativity \cite{1}. The latter classification is of course fundamental
but at the same time equivocally general, in the sense that spacetimes
belonging
to one and the same Bianchi type may differ profoundly in other important
aspects. In other words, while the Bianchi assignments actually classify
{\em types} of 3D homogeneous spaces (essentially
their $G_3$ isometry groups with no regard to the 4D metric),
we seek a classification of the 4D spacetimes themselves.
With the first hints comming from the class of solutions
presented in \cite{6}, it is by now evident that, in the context of string
cosmology, the Bianchi-type assignments
may be ramified in the above sense to yield a comprehensive and illuminating
classification
of all possible HSBs. With this paper we aim at presenting such
a classification explicitly.

Of the total 576 HSBs classified,
quite a few are presently known and
we will try to fascilitate
their identification in the existing literature.
In the following
section we have gathered some general definitions and facts needed for
the presentation of our main results in section 3,
further discussed in section 4. The latter also contains a tabulation of
our brief review on the subclass of `diagonal' HSBs. The quotation marks
serve as a reminder that the diagonality of the metric
just been refered to holds in the invariant non-holonomic basis $\{\sigma^i\}$
(to be explicitly discussed shortly).

\section{Preliminaries on the parameters of classification}

We want to classify all 4D spacetimes with metrics of the form
\begin{equation}
ds^2=-dt^2+g_{ij}(t)\sigma^i\sigma^j, \label{met}
\end{equation}
as part of a background solution which satisfies at least the lowest-order
string beta-function equations
for conformal invariance.
To fix notation and conventions used, we recall that these equations
can be derived from the effective action \cite{2}
\begin{equation}
S_{eff}=\int d^4x
\sqrt{-g}e^{\phi}(R-\frac{1}{12}
H_{\mu\nu\rho}H^{\mu\nu\rho}+\partial_{\mu}\phi\partial^{\mu}\phi-\Lambda)
\label{b4'}.
\end{equation}
In this  (so-called `sigma-') conformal frame they are
\begin{eqnarray}
R_{\mu\nu}-\frac{1}{4}H_{\mu\nu}^2-\nabla_\mu\nabla_\nu\phi&=&0,
\label{b1}\\
\nabla^2(e^\phi H_{\mu\nu\lambda})&=&0, \label{b2} \\
-R+\frac{1}{12}H^2+2\nabla^2
\phi+(\partial_\mu\phi)^2+\Lambda&=&0,\label{b3}
\end{eqnarray}
where $\Lambda$ is the cosmological constant emerging as a result of a
non-vanishing
central charge deficit in the original theory.
In addition to the gravitational field $g_{\mu\nu}$, these expressions also
involve
the two other fundamental bosons present in the effective string action,
namely the dilaton $\phi$ and,
in the contractions $H_{\mu\nu}^2=H_{\mu\kappa\lambda}
{H_{\nu}}^{\kappa\lambda}
\, , H^2=H_{\mu\nu\lambda}H^{\mu\nu\lambda}$, the totally antisymmetric
field strenght $H_{\mu\nu\lambda}$. The latter, which may be equivalently
viewed here as a closed 3-form, is defined in terms of the potential
$B_{\mu\nu}$ as
\begin{equation}
H_{\mu\nu\rho}=\partial_\mu
B_{\nu\rho}+\partial\rho B_{\mu\nu}+\partial_{\nu} B_{\rho\mu} \label{H}.
\end{equation}

The expression (\ref{met}) adopted above gives in fact
the most general {\em synchronous} metric possible. A universal time t
is well defined in such manifolds, in which the $t=const.$ hypersurfaces
of simultaneity actually coincide with the $\Sigma^3$ hypersurfaces of
homogeneity, to be defined shortly.
The metric coefficients, expressed in (\ref{met}) in terms of the $3\times3$
matrix $g_{ij}$,
can be functions of t  only
and $\{\sigma^i,\, i=1,2,3\}$ is the already mentioned basis of 1-forms,
invariant under the left action of a 3-parameter group of motions $G_3$.
In view of their fundamental importance, these invariant forms
will be precisely defined, together with
the class of HSBs. The latter consists of all 4-dimensional spacetime
manifolds $M^4$ which admit an r-parameter group of isometries $G_r$ whose
orbits in $M^4$ are 3-dimensional space-like hypersurfaces.
As in conventional general relativity \cite{1}, these
are precisely the hypersurfaces of homogeneity  $\Sigma^3$ on
which a $G_3$ subgroup of $G_r$ acts transitively. In view of the
dimensionality of $M^4$,
we can only have $3\leq r\leq 6$. Any
remaining symmetry (and the corresponding independent Killing vectors)
must generate the (r-3)-dimensional
isotropy subgroup of $G_r$. Since there is no
two-dimensional rotation group, (r-3) can only have the values $0,1,3$,
the last one associated with the maximal FRW-type of symmetry. The
action of $G_3$ is generaly transitive on its orbits.
There is only one exception,
realized in the Kantowski-Sachs type of metric, in which $G_3$
is multiply transitive, actually acting on 2-dimensional space-like surfaces of
maximal symmetry \cite{1}.
This is a marginal case (which is not to say it is unimportant!),
which will be counted in our classification but it will not be considerd any
further in the present work.
For the main and typical case,
all  possible isometry  groups $G_3$ of  the metric (\ref{met})
are known and have been fully classified
in nine Bianchi types, each identified
by the corresponding set of group-structure constants $C^i_{jk}$. The latter
also define
(essentially uniquely) the invariant $\sigma^i$ 1-forms by the relation
\begin{equation}
d\sigma^i=-\frac{1}{2}C^i_{jk}\sigma^j\wedge\sigma^k
\end{equation}
Equivalently, the dual relation
\begin{equation}
[\xi_i,\xi_j]=-C^k_{ij}\xi_k,
\end{equation}
involves a set of three independent Killing vectors ${\xi_i}$ which form a
basis dual to $\{\sigma^i\}$.
All Bianchi-types have been further characterized in the literature \cite{1} as
being of class A if their
adjoint representation  is traceless, otherwise they are of class B.

Classification parameters are in priciple expected
to come from the cosmological constant $\Lambda$,
the metric (\ref{met}),
as well as the dilaton and $H$ fields.
The last two must also respect the $G_3$-isometries, namely their Lie
derivatives wrt any
Killing vector formed by the $\{\xi_i\}$ basis must vanish. Equivalently,
the dilaton field must be a constant on $\Sigma^3$ and the H-field, when as a
3-form is projected in $\Sigma^3$, must
equal a constant times the invariant 3-form in $\Sigma^3$.
Thus, when viewed in $M^4$, the dilaton $\phi$ can only be a function of the
time t.
On the other hand, the dual $\star H$ of $H$ expressed in the same basis as
(\ref{met}) must be of the form
\begin{equation}
\star H=H^{\ast}_0(t)dt + H^{\ast}_i(t)\sigma^i.\label{dualh}
\end{equation}
namely with components $H^{\ast}_\mu$ at most functions of t
in the $\{\sigma^i\}$ basis.

\section{The classification of 4D HSBs}
We are now in a position to proceed with a classification of all possible HSBs
in 4D.
The presence of a cosmological constant
contributes with a factor of 2 in the
number of possible HSBs, in other words we have two major
subclasses corresponding the $\Lambda=0$ and $\Lambda\neq 0$ cases.
They will both be counted but all subsequent discussion will be restricted to
the
$\Lambda=0$ case. To examine the r\^{o}le of the fundamental fields,
we firstly observe that the dilaton,
been a scalar and, in principle, an ever-present one,
can hardly be expected to have any effect on
the classification. To see explicitly that this is indeed the case, one must
examine
the contribution of the $\phi$ field in (\ref{b1}), which will be given
shortly.
We note, however that the result of this calculation
is more illuminating if presented {\em not\/} in the frame in which (\ref{met})
is
expressed but rather in an orthonormal frame $\{\omega^\mu\}$.
The latter may always be introduced (see, eg, \cite{6}) so that (\ref{met})
becomes
\begin{equation}
ds^2=\eta_{\mu\nu}\omega^\mu\omega^\nu, \label{metd}
\end{equation}
where $\eta_{\mu\nu}$ stands for the Minkowski signature $(-1,1,1,1)$. In the
so chosen
$\{\omega^\mu\}$ basis and with a dot for $d/dt$ we find
\begin{equation}
\nabla_\mu\nabla_\nu\phi=
\left(\begin{array}{cc}
\ddot{\phi}&0\\0&\gamma_{oij}\dot{\phi}\end{array}\right) , \label{fiem}
\end{equation}
where $\gamma_{0ij}$ are the (non-holonomic) connection coefficients of
(\ref{metd}).
Considered as a $3\times3$ matrix, $\gamma_{0ij}$ is diagonal whenever $g_{ij}$
in (\ref{met}) is diagonal. However, even in the latter case, the Ricci tensor
can have non-vanishing off-diagonal components in the orthonormal frame. Thus
the form of the contribution of the dilaton in the gravitational
part of the field equations is always superceeded by the presence of purely
gravitational contributions comming from $g_{ij}$.

Turning to the gravitational field, the r\^{o}le of $g_{ij}$ is  more
transparent if examined in the same orthonormal frame, as one might have
expected.
We have just mentioned that even when diagonal, the
$g_{ij}$ matrix may give off-diagonal $(i\neq j)$ equations in the set
(\ref{b1}) in the $\{\omega_\mu\}$ frame.
That happens in the case of Bianchi types which are of class B. However, in
that case, the off-diagonal
equations are just {\em constraints\/} on the diagonal ones \cite{6}.
In the case of a non-diagonal $g_{ij}$, obviously with up to 3 independent
functions $g_{ij}(t)$, there will be
as many independent off-diagonal equations. The rest of the off-diagonal
equations
in the set (\ref{b1}) will retain their constraint-like character.
It is thus clear that there exist two major subclasses corresponding
to  $g_{ij}$ been diagonal and non-diagonal. It
should be recalled, however, that diagonality as well as dependence
 on the coordinates are frame-depended properties, so that
the above picture would appear completely different if,
 instead of the
{\em non-holonomic\/} frames $\{\sigma^\mu\}$ and $\{\omega^\mu\}$,
 one employed conventional ones. We will not use the latter here but
one could explicitly find holonomic-coordinate expressions
by utilizing the appropriate entries supplied by the Table in the last section.
Similar arguments and likewise dependence on t as well as on the spatial
coordinates
hold for any other $G_3$-invariant quantity.
This is in particular true for the components of $H$ or its dual $\star H$, the
contribution of which we will
investigate next.

One can easily compute the contribution of H in
(\ref{b1}) in the $\{\omega^\mu\}$ frame and subsequently express
everything in terms of the dual $\star H$.
As a result of this straightforward calculation one
distinguishes
four categories of H-field configurations which we may denote in order of
increasing complexity
 as $(0)$, $(\uparrow)$,$(\rightarrow)$,$(\nearrow)$.
Of theese, the first one corresponds to the trivial case of vanishing H. Next
is the case of H fields such that
their dual $\star H$ (seen as a vector) is orthogonal
to the (pictured as `horizontal') hypersurfaces of homogeneity $\Sigma^3$.
Equivalently, a $(\uparrow)$ configuration
is such that $H^{\ast}_i=0$ in (\ref{dualh}).
Complementary to that is the `transverse' case of a $\star H$ lying entirely
whithin the $\Sigma^3$ hypersurfaces,
namely with $H^{\ast}_0=0$, suggestively denoted as $(\rightarrow)$.
H fields `tilted' between the two extremes corespond to the general
(\ref{dualh}) case, represented
by the tilted arrow $(\nearrow)$. Of all four configurations only the first
two have been deployed in the literature on HSBs until now.
They will all be further discussed in the last section.

The classification of HSBs may now proceed as follows.
Without the need to any explicit reference to the values of
$\Lambda$ and $\phi$, each classified HSB may be generically codified as
$X(n,d,a)$.
In that acronym, X will generally denote
the Bianchi type of $G_3$, so that X essentially takes
 the values $I,II,\ldots ,IX$. It should be noted, however, that the
multitude of the distinct values taken by X is in fact 12: nine
for the conventional Bianchi types, plus two for a necessary refinement
(concerning types $VI_h$ and $VII_h$, as it will be explicitly seen
in the sequel) plus one for the Kantowski-Sachs case.
The argument n specifies the isotropy group and takes three distinct values:
it equals 3 for $SO(3)$ (the case of complete
isotropy), it equals 2 for $SO(2)$ (only two principal directions are
equivalent) and
it is omitted altogether if there is no isotropy group. The next argument takes
two distinct values: it
will be present only when the $g_{ij}$ matrix in (\ref{met}) is diagonal.
The last argument (a) takes the
four values $(0)$, $(\uparrow)$,$(\rightarrow)$,$(\nearrow)$, just discussed.
Thus the
product of the multiplicities of the three aguments is $3\times 2\times 4$=24.
Taking also
into account the multiplicity of X, as well as the
factor of 2 coming from the cosmological constant as mentioned,
we arrive at the earlier quoted number of $24^2=576$ HSBs, including
a subclass of FRW or FRW-like models which will be counted in the
next section. In the rest of this section we will briefly discuss the X(d)
backgrounds, namely
the subclass of `diagonal' metrics. As mentioned, an effort will be made
to fascilitate the identification of HSBs which are already known, however
without aiming at what
should rather be the objective of a review article. These results will be
further
discussed in the last section.

\vspace{1cm}
{\bf Type I\@.\/}
Spacetimes of this type admit an isotropy limit which contains
all possible flat ($k=0$) FRW-type of HSBs. Until recently, only the $I(3d0)$
model
and its generalizations $I(3d\uparrow)$ and the Kasner-like $I(d0)$
had been given\cite{3},\cite{4},\cite{5}.
They are all reproduced as special cases of the $I(d\uparrow)$ HSB given in
\cite{6}.
Very recently, the $I(d\rightarrow)$ has been found, together with the result
that
there exist no $I(d\nearrow)$ solution \cite{7}.

{\bf Type II\@.\/}
The fully anisotropic $II(d\uparrow)$ was given in \cite{6}, generalizing the
$II(d0)$ found in \cite{5}.
 For the $II(d\rightarrow)$ and $II(d\nearrow)$ cases the same hold as for type
I.

{\bf Type III\@.\/}
The general anisotropic solution, namely $III(d\nearrow)$, exists and reduces
to the $III(d\uparrow)$ found in\cite{6} which in turn reduces to the $III(d0)$
found in \cite{5}.
However, there also exists a general $III(d\rightarrow)$ HSB which cannot be
reached as a limit of
the mentioned $III(d\nearrow)$ (cf. \cite{7}).

{\bf Type IV\@.\/}
There exist no $IV(d)$ HSBs, as all diagonal solutions are singular everywhere
\cite{6},\cite{7}.

{\bf Type V\@.\/}
For HSBs of this type
the isotropy limit attainable
gives rise to the open ($k=-1$) FRW-type
spacecetimes. Such are the $V(3d0)$ and $V(3d\uparrow)$ cases
found in \cite{3}, \cite{4}, respectively. The first has been
generalized by the $V(d0)$ in \cite{5}, while all are special cases of
$V(d\uparrow)$ HSB found in \cite{6}.
The later is in turn the obvious limit of the general `diagonal' case, namely
$V(d\nearrow)$.
Curiously enough, there exists no $V(d\rightarrow)$ solution \cite{7}.

{\bf Type $VI_h$\@.\/}
The $G_3$ involved in this type of models actually form a
continuous 1-parameter family of groups parametrized by  $h$,
with the values  $h\neq 0,1$ typically excluded as giving rise to
Bianchi types III and V respectively (all these types are of class B).
The results of the previous case are generally valid here as well, except of
course for the isotropy limit.
The real exception is with the $h=-1$ case, which must be and is discussed
separately next.

{\bf Type $VI_{-1}$\@.\/}
HSBs in this case are generally {\em not} obtained at the $h=-1$ limit from
solutions of the
previous type. The general `diagonal' solution, namely $VI_{-1}(d\nearrow)$
has been recently found. It generalizes
the $VI_{-1}(d\uparrow)$ in \cite{6}, and also gives a non-trivial
$VI_{-1}(d\rightarrow)$ limit \cite{7}.

{\bf Type $VII_h$\@.\/}
In this case, which also involves a 1-parameter group $G_3$ (here with $h^2\leq
4$),
all solutions are singular  everywhere, namely there exist no $VII_h(d)$ HSBs
unless $h=0$.
The latter case (which exceptionally involves
space-times of  class A) must be and is considered separately next.

{\bf Type $VII_0$\@.\/}
There exists the expected $VII_0(d\uparrow)$  with isotropy
limits $VII_0(2d\uparrow)$ and $VII_0(3d\uparrow)$ as in the Type-I case
\cite{6}.
As recently found, there also exists the $VII_0(d\rightarrow)$ but no
$VII_0(d\nearrow)$ solution \cite{7}.

{\bf Type VIII\@.\/}
The $VIII(2d\uparrow)$ has been found in \cite{6}, while neither
$VIII(d\nearrow)$ nor $VIII(d\rightarrow)$ exist \cite{7}.

{\bf Type IX\@.\/}
The $IX(2d\uparrow)$ case ( namely the analogue of the well-known Taub
metric to which it reduces) has been recently found \cite{6}.
The complete isotropy limit, namely $IX(3d\uparrow)$,
exists and reproduces the closed $(k=1)$ FRW-type of solutions found in
\cite{3},\cite{4}.
There are no other `diagonal' Bianchi-type IX
HSBs \cite{7}
except for the elusive $IX(d\uparrow)$
(escaping us just like its Mixmaster countrpart!).

\section{Conclusions}

\newcommand{\e}{$\boldmath\exists$}
\newcommand{\n}{$\not\!\exists$}
\newcommand{\ep}{$\exists^\ast$}
\newcommand{\ri}{$\;\Rightarrow$}

A classification of all possible 4D HSBs, codified as X(n,d,a), has been
presented.
It is based on the known Bianchi classification of the isometry groups $G_3$
which generate all homogeneous 3D spaces, so that X takes the values
I,II,\ldots,IX etc (roughly one for
each of the Bianchi types plus one for the Katowski-Sachs class of metrics).
Of the the arguments n,d,a one or more may
be omitted or take values as follows. The integer n, which identifies the
isotropy group,
takes the value 3 for SO(3), 2 for SO(2) and it is omitted in the case of
complete anisotropy. The argument d is omitted only when $g_{ij}$ in
(\ref{met}) is
{\em non-}diagonal in the synchronous frame of the invariant $\{\sigma^i\}$
basis.
The third argument, zero in the trivial case of an identically vanishing H
field,
indicates the orientation of $\star H$ wrt the hypersurfaces of homogeneity
$\Sigma^3$.

The quoted number of $24^2=576$ HSBs which the classification sees as distinct,
includes a subclass of 192 backgrounds which are either
FRW or
could asymptotically attain full SO(3) isometry.
They can be easily counted
because they may only descend from type I(and VII) for the flat (k=0) case, or
from type V for the open (k=-1), or from type IX for the closed (k=1) models
(so their multitude is $2\times4\times(3\times2\times4)$). Some of the HSBs
which the classification sees as distinct are in fact special cases of higher
symmetry, while others cannot
be considered as practically realizable  because their manifolds are singular
everywhere.
Such is the case for those denoted by \n $\,\,$in the the Table below. The
latter gives a summary of the `diagonal'
types discussed in the previous section, with the arrow \ri $\,\,$  pointing
towards
special sub-cases, mostly  the just-mentioned higher-symmetry limits.

The most rudimentary physical aspects of the parameters of classification have
already been discussed, particularly concerning the isometry and isotropy
groups. Remaining to be
so discussed is the parameter which specifies the orientation of any given
H-field configuration.
That orientation may be better visualized
in terms of an associated {\em congruence} of flow lines to which $\star H$ is
tangent.
Such a congruence will cross the family of $\Sigma^3$ (as the latter evolves in
time) orthogonally $(\uparrow)$,
or it may be tilted $(\nearrow)$, possibly to the limit of
lying entirely within those hypersurfaces $(\rightarrow)$.
It will sufice to examine the Bianchi-type V case, for which we copy from our
Table the sequence
\begin{equation}
\cdots \,\, V(d\nearrow)\,\,\Rightarrow \,\,V(d\uparrow)\,\,\Rightarrow \,\,
\cdots \,\,\Rightarrow \,\, V(3d\uparrow).
\end{equation}
In the $V(d\nearrow)$ model the $\star H$ congruence will have general
kinematics, namely it will
exhibit expansion, shear (anisotropy) and, in principle,  vorticity as well. In
the next more
symmetric case $V(d\uparrow)$ vorticity  cannot be sustained, while in the
fully symmetric $V(3d\uparrow)$
(namely the open FRW model) only expansion has survived.

We have already mentioned that, in the context of HSBs classified here, one
might
appropriately examine certain deeper aspects of string theory. These are mostly
related
to the construction of realistic CFTs or the retention of conformal invariance
and other symmetries under duality transformations (cf, eg, \cite{5},\cite{8}
and refs cited therein).
Here we will briefly discuss the behavior of HSBs under abelian target space
duality.
As first noted in \cite{6}, $X(d\uparrow)$ backgrounds produce duals with
metrics in the same class,
exept fot the cases marked as \ep $\,\,$ in the table. These, apparently
loosing most of
their symmmetry, give
rise to non-homogeneous backgrounds. However, even the relatively much simpler
$X(d\rightarrow)$
class may produce highly non-trivial duals (inhomogeneous and without universal
time
in view of the emergence of $g_{ti}$ components). For the general case, it
appears
that $X(\nearrow)$ may generate backgrounds so general that they would be
virtually
ureachable by any other approach.

\vspace{1cm}
I would like to thank E. Kiritsis and C. Kounnas for discussions.
\newpage
\leftmargin=0cm
\begin{center}
{\bf Classification of `diagonal' 4D HSBs}
\end{center}

\begin{center}
\begin{tabular}{|c|c|c|c|c|c|c|} \hline
Coordinate basis & X & $d\rightarrow$ & $d\nearrow$ & $d\uparrow$ &
$2d\uparrow$ & $3d\uparrow$ \\
\hline \hline

 & & & & & & \\
$\sigma^i=dx^i$&I&\e &\n &\e \ri &\e \ri &\e \\
 & & & & & & \\ \hline
$\sigma^1=dx^2-x^1dx^3$& & & & & & \\
$\sigma^2=dx^3$&II&\e &\n &\e &\n &\n \\
$\sigma^3=dx^1$& & & & & &\\ \hline
$\sigma^1=dx^1$& & & & & & \\
$\sigma^2=dx^2$&III&\e &\e \ri &\e &\n &\n \\
$\sigma^3=e^{x^1}dx^3$& & & & &  & \\ \hline
$\sigma^1=e^{-x^1}dx^2-x^1e^{-x^1}dx^3$& & & & & &\\
$\sigma^2=e^{-x^1}dx^3$&IV&\n&\n&\n&\n&\n\\
$\sigma^3=dx^1$& & & & & & \\ \hline
$\sigma^1=dx^1$& &  & & & &\\
$\sigma^2=e^{x^1}dx^2$&V&\n &\e \ri &\e \ri &\e \ri &\e \\
$\sigma^3=e^{x^1}dx^3$&  & & & & & \\ \hline
$\sigma^1=dx^1$& & & & & & \\
$\sigma^2=
e^{hx^1}dx^2$&$VI_h$&\n &\e \ri &\e &\n &\n \\
$\sigma^3=e^{x^1}dx^3$& & & & & &\\ \cline{2-7}
 & & & & & & \\
$(h=-1)$&$VI_{-1}$&\e &$\Leftarrow$\e \ri &\e &\n& \n \\
 & & & & & & \\ \hline
$\sigma^1=(A-kB)dx^2-Bdx^3$& & & & & &\\
$\sigma^2=Bdx^2+(A+kB)dx^3$&$VII_h$&\n &\n &\n &\n &\n \\
$\sigma^3=dx^1$& & & & & & \\ \cline{2-7}
 & & & & & & \\
$(h=0)$&$VII_0$&\ep &\n &\ep \ri &\e \ri &\e \\
 & & & & & & \\ \hline
$\sigma^1=dx^1
+((x^1)^2-1)dx^2+Z_1dx^3$& & & & & & \\
$\sigma^2=dx^1
+((x^1)^2+1)dx^2+Z_2dx^3$&$VIII$&\n &\n &\ep \ri &\e &\n \\
$\sigma^3=2x^1dx^2+Z_3dx^3$& & & & & & \\ \hline
$\sigma^1=-\sin x^3dx^1+\sin x^1\cos x^3dx^2$& & & & & & \\
$\sigma^2=\cos x^3dx^1+\sin x^1\sin x^3dx^2$&$IX$&\n &\n &\ep \ri &\e \ri &\e
\\
$\sigma^3=\cos x^1dx^2+dx^3$& & & & & &\\ \hline
\end{tabular}
\end{center}
\vspace{.2cm}

{\em notes\/}:

\ep $\,\,$ indicates that the solution exists but it has not been found in
closed form.

The last column gives the FRW limits.

 $ A=e^{-kx^1}\cos(qx^1)\,,\, B=-\frac{1}{q}e^{-kx^1}
\sin(qx^1)\,,\,\, \, \, \,
[k=\frac{h}{2}, \, q=\sqrt{1-k^2}]$

$Z_1=x^1+x^2-x^2(x^1)^2$,\, \, \,
$Z_2=x^1-x^2-x^2(x^1)^2$,\,\, \,
$Z_3=1-2x^1x^2$.

\end{document}